\documentclass[aps,pre,twocolumn,a4paper,endfloats]{revtex4}

\usepackage{graphicx}
\usepackage{bm}
\usepackage{amssymb}
\usepackage{latexsym}
\usepackage{amsmath}

\begin{document}

\title{Implicit and explicit solvent models for the simulation of a single
polymer chain in solution: Lattice Boltzmann vs Brownian dynamics}

\date{\today}

\author{Tri T. Pham}
\affiliation{Department of Chemical Engineering, Monash University,
Clayton, VIC~3800, Australia} \affiliation{Max Planck Institute for
Polymer Research, Ackermannweg 10, D-55128 Mainz, Germany}

\author{Ulf D. Schiller}
\affiliation{Max Planck Institute for Polymer Research, Ackermannweg
10, D-55128 Mainz, Germany} \affiliation{Chemical Engineering Dept.,
University of Florida, Gainesville, FL 32611-6005, USA}

\author{J. Ravi Prakash}
\affiliation{Department of Chemical Engineering, Monash
University, Clayton, VIC~3800, Australia}

\author{Burkhard D\"{u}nweg}
\affiliation{Max Planck Institute for Polymer Research, Ackermannweg
10, D-55128 Mainz, Germany}

\begin{abstract}

  We present a comparative study of two computer simulation methods to
  obtain static and dynamic properties of dilute polymer solutions.
  The first approach is a recently established hybrid algorithm based
  upon dissipative coupling between Molecular Dynamics and lattice
  Boltzmann (LB), while the second is standard Brownian Dynamics (BD)
  with fluctuating hydrodynamic interactions. Applying these methods
  to the same physical system (a single polymer chain in a good
  solvent in thermal equilibrium) allows us to draw a detailed and
  quantitative comparison in terms of both accuracy and efficiency. It
  is found that the static conformations of the LB model are distorted
  when the box length $L$ is too small compared to the chain size.
  Furthermore, some dynamic properties of the LB model are subject to
  an $L^{-1}$ finite size effect, while the BD model directly
  reproduces the asymptotic $L \to \infty$ behavior. Apart from these
  finite size effects, it is also found that in order to obtain the
  correct dynamic properties for the LB simulations, it is crucial to
  properly thermalize all the kinetic modes. Only in this case, the
  results are in excellent agreement with each other, as expected.
  Moreover, Brownian Dynamics is found to be much more efficient than
  lattice Boltzmann as long as the degree of polymerization is not
  excessively large.

\end{abstract}

\maketitle

\section{\label{sec:Intro}Introduction}

The rich variety of conformations which leads to many different
intrinsic properties of polymer solutions has continuously drawn
considerable interest in soft matter research. Computer modeling is
increasingly being used as an integral part of theoretical study, in
order to both test existing theories and to trigger the development of
new concepts. Furthermore, computer simulations have also become an
essential tool in materials research, especially for predicting and
understanding the behavior of complex systems, where a complete theory
is not available. It has been proven to be an effective and
inexpensive way to study these systems. In order to observe
large-scale properties, it is crucial to reduce the computational cost
by coarse-graining the details of the atomic structure. This is
particularly true for polymer systems and studies of their universal
static and dynamic properties \cite{deGennes79,DoiEdwards86}. In this
context, using a conventional bead-spring chain model to represent a
polymer molecule in Molecular Dynamics (MD) simulations is usually
sufficient \cite{KremerGrest90,PierleoniRyckaert92,SmithRapaport92,%
  DuenwegKremer93,Binder95}. In the case of dilute and semidilute
polymer solutions, a correct model also needs to take into account the
effect of solvent molecules. This effect is two-fold: On the one hand,
the good solvent quality results in swelling of the random coil; on
the other, the solvent-mediated long-range dynamic correlations
between different segments of the chain, known as hydrodynamic
interactions (HI), significantly influence the dynamical behavior
\cite{deGennes79,DoiEdwards86,SuntharPrakash06}. In the present
methodological study, we focus on the dilute regime which is
theoretically most thoroughly understood.

In order to capture hydrodynamic interactions in MD simulations, the
solvent particles need to be incorporated explicitly. Typically, the
number of solvent particles required for such a model is of the order
of thousands even for a short chain. Although such studies are
feasible \cite{DuenwegKremer93}, they are rather inefficient, for this
reason. Therefore, a more coarse-grained description of the solvent is
highly desirable. Two complementary approaches have been developed to
do this. ``Mesoscopic'' methods keep the solvent degrees of freedom,
but describe them in a simplified fashion. These include Dissipative
Particle Dynamics (DPD)
\cite{HoogerbruggeKoelman92,Schlijperetal95,EspanolWarren95,%
Marshetal97,GrootWarren97,Espanol98,Pagonabarragaetal98},
Multi--Particle Collision Dynamics (MPCD)
\cite{MalevanetsKapral99,LeeKapral06,Gompper09}, and lattice
Boltzmann (LB)
\cite{Benzietal92,Ladd94a,Ladd94b,ChenDoolen98,Succi01,%
LaddVerberg01,DuenwegLadd09,AhlrichsDuenweg98,AhlrichsDuenweg99,%
Adhikarietal05,Duenwegetal07}.
These approaches are typically one to two orders of magnitude faster
than MD \cite{AhlrichsDuenweg99}. Conversely, Brownian Dynamics (BD)
simulations \cite{ErmakMcCammon78,Jendrejacketal02,LiuDuenweg03,%
Prabhakaretal04b,SuntharPrakash05} remove the solvent degrees of
freedom completely, and take their effect into account via
non-trivial long-range dynamic correlations in the stochastic
displacements. This is possible due to the time scale separation
between the fast solvent motion and the slow conformational polymer
degrees of freedom. Since the number of degrees of freedom is
reduced drastically, the method has the potential to save CPU time
by additional several orders of magnitude, in particular in the
dilute limit. However, a simple implementation of the correlations
\cite{ErmakMcCammon78} leads to an algorithm which scales like
$O(N^3)$, where $N$ is the number of Brownian particles, and
therefore becomes infeasible as soon as $N$ exceeds a few hundred
\cite{LiuDuenweg03}. It is therefore very important to treat HI by
means of Fixman's algorithm \cite{Fixman86} (scaling roughly as
$O(N^{2.25})$), which we do in the present study. The recently
introduced method by Geyer and Winter \cite{GeyerWinter08}
would reduce the necessary CPU effort by roughly one additional
order of magnitude for typical chain lengths, and exhibit
a more favorable scaling ($O(N^{2})$). However, it
is based upon an inexact approximation of the hydrodynamic
correlations that cannot be improved systematically (in
contrast to Fixman's method). For this reason, this algorithm
was not implemented.

None of these approaches is sufficient to reach $N \sim 10^3 \ldots
10^4$; this latter goal is only attainable by the implementation of
very recent ``superfast'' BD algorithms based upon Fast Fourier
Transforms \cite{BanchioBrady03,Saintillanetal05,%
  Hernandez-Ortizetal07}. These latter algorithms scale as $N^{1+x}
\log N$, where the exponent $x$ depends on the details of the
underlying physics, and is usually substantially smaller than unity.
These methods require the study of a confined system, and hence are
not used in the present study.

While the advantages and disadvantages of the methods are well-known
in general terms (and have resulted in differing methodological
preferences in different groups of researchers), not much is known
\emph{quantitatively} in terms of a clear comparison of computational
efficiency. The present paper aims at partly filling this gap.

Recently, one of the present authors
\cite{AhlrichsDuenweg98,AhlrichsDuenweg99} has proposed a new
mesoscopic method for simulating polymer-solvent systems. The
solvent is represented by a fluid on a grid, simulated via the
lattice Boltzmann approach, while the motion of the polymer chain is
governed by a continuous MD model. The two parts are coupled by a
simple dissipative force. The lattice Boltzmann (LB) method was
originally developed to simulate hydrodynamics on a grid
\cite{Benzietal92,ChenDoolen98}. It has been shown that the LB
method is a fast and effective method for simulating fluid flows,
which has the same speed and accuracy as other Navier-Stokes solvers
\cite{Benzietal92,Ladd94a,Martinezetal94,Ladd94b}. Ladd
\cite{Ladd94a,Ladd94b} successfully applied the LB method to
colloidal systems (originally with a \emph{conservative} coupling)
and showed that the CPU cost scales linearly with the number of
particles. Moreover, he showed how fluctuations can be incorporated
into the LB model, which is essential in order to investigate
Brownian motion \cite{Ladd94a}. This procedure has recently been
refined and improved \cite{Adhikarietal05,Duenwegetal07}. The
dissipative coupling method
\cite{AhlrichsDuenweg98,AhlrichsDuenweg99} was thoroughly tested by
applying it to a single polymer chain in solution, for which the
data of a previous MD simulation \cite{DuenwegKremer93} were
available, and whose parameters were used as an input for the
mesoscopic model.

In this work, we study the dynamics of a single chain in a solvent to
compare the predictions of the explicit solvent model via the LB
method with the predictions of the implicit solvent model by BD
simulations. Many recent simulation studies have investigated very
similar systems, using various mesoscopic solvent models like MPCD
\cite{mussawisade}, DPD \cite{Jiangetal07}, direct solution of the
Navier-Stokes equation \cite{giupponi}, and smoothed DPD
\cite{litvinov}. These studies were mainly done in order to test and
verify the validity of the chosen mesoscopic simulation approach.
Meanwhile it is fair to say that a single chain in solution is a
standard benchmark test bed for mesoscopic simulation methods, the
necessary condition for passing being the correct reproduction of the
Zimm scaling laws within the limitations of finite chain length and
finite solvent volume. The present study, however, aims directly at a
\emph{quantititave comparison} between two different methods, and
therefore it is crucial that that the underlying polymer model is
exactly identical for both methods. This is the reason why the data of
Refs. \cite{mussawisade,Jiangetal07,giupponi,litvinov}, though all
exhibiting essentially the same physics, are of no direct relevance
for the present investigation, since all of them employ slightly
different polymer models, and simulate solvents with somewhat
different viscosities. One very recent study by Ladd et
al. \cite{laddcomparison} has done a rather similar comparison between
LB and BD, and arrived at similar conclusions; the present paper
should be viewed as a complementary study that puts some more emphasis
on the issue of computational efficiency.

After introducing the models, we show how to map the input parameters
of the hybrid model onto the input values of the BD model to directly
compare the predicted quantities (Sec.~\ref{sec:Models}).
Section~\ref{sec:Results} then confirms the expected physical
equivalence of the two approaches in terms of comparing static and
dynamic data. Furthermore, this section also presents our comparison
on the numerical cost or benchmark data for both methods. Finally, in
Sec.~\ref{sec:Conclusions} we summarize the results and give our
conclusions.

\section{\label{sec:Models}
Molecular model and simulation methods}

\subsection{Molecular model}

In this work, a polymer molecule is represented by a conventional
bead-spring chain model, which consists of $N$ beads that are
connected via $N-1$ finitely extensible nonlinear elastic (FENE)
massless springs. The Lennard-Jones potential, which acts between
all monomers, is used to model the excluded volume (EV) effect. The
two potentials $V_\text{FENE}$ and $V_\text{LJ}$ are given by the
expressions
\begin{eqnarray}
V_{\text{FENE}} & = & - \frac{k_{\text{FENE}}R^2_0}{2}
      \ln \left(1 - \left( \frac{r}{R_0} \right)^2 \right) , \\
V_{\text{LJ}} & = & 4 \epsilon \left( \frac{\sigma^{12}}{r^{12}}
             - \frac{\sigma^6}{r^6} + \frac{1}{4} \right) ,
\quad r \leq 2^{1/6} \sigma ,
\end{eqnarray}
where $r$ is the bead-bead distance, $k_\text{FENE}$ is the spring
constant and $R_0$ the maximum extension of the bond. $\epsilon$ and
$\sigma$ are the energy and length parameters of the Lennard-Jones
potential, respectively.

\subsection{The lattice Boltzmann (LB) method}

In this method, the evolution of the LB variables $n_i$ is governed
by the following lattice Boltzmann equation
\cite{Benzietal92,Succi01,DuenwegLadd09}:
\begin{eqnarray}
&   & n_i(\mathbf{r}+\mathbf{c}_i\Delta\tau,t+\Delta\tau)
  =   n_i(\mathbf{r},t) \\
& + & \sum_{j=1}^{b} L_{ij} \left( n_j (\mathbf{r},t)
      - n_j^{\text{eq}}(\rho,\mathbf{u}) \right) + n'_i(\mathbf{r},t) .
\nonumber
\end{eqnarray}
The variable $n_i(\mathbf{r},t)$ is the (partial) fluid mass density
at grid site $\mathbf{r}$ at time $t$, corresponding to the discrete
velocity $\mathbf{c}_i$. $\Delta\tau$ is the time step, and the
lattice spacing is denoted by $a$. The small set of velocities
$\mathbf{c}_i$ ($i=1, \ldots, b$, where the value of $b$ depends on
the details of the model) is chosen such that $\mathbf{c}_i
\Delta\tau$ is a vector leading to the $i$th neighbor on the grid.
$L_{ij}$ is a collision operator for dissipation due to fluid particle
collisions, such that the populations always relax toward the local
pseudo-equilibrium distribution $n_j^{\text{eq}}$ that depends on the
local hydrodynamic variables $\rho = \sum_i n_i$ (the total mass
density) and $\mathbf{u} = \sum_i n_i \mathbf{c}_i / \sum_i n_i$ (the
local flow velocity). The collision process is constructed in such a
way that it conserves both $\rho$ and $\mathbf{u}$.
$n'_i(\mathbf{r},t)$ is the stochastic term, which is essential in
order to simulate thermal fluctuations that drive Brownian motion.

The local pseudo-equilibrium distribution can be represented as a
second-order expansion of the Maxwell-Boltzmann distribution, given
by \cite{Succi01}
\begin{equation}
n_i^{\text{eq}}(\rho,\mathbf{u}) = \rho w_{c_i} \left(
1 + \frac{\mathbf{c}_i\cdot\mathbf{u}}{c^2_\text{s}}
+ \frac{(\mathbf{c}_i\cdot\mathbf{u})^2}{2c^4_\text{s}}
- \frac{u^2}{2c^2_\text{s}} \right) ,
\end{equation}
where $w_{c_i}$ are a set of weight factors, which depend on the
sublattice $i$ (i.~e. the magnitude of $\mathbf{c}_i$) and
$c_\text{s}=\sqrt{1/3}(a/\Delta\tau)$ is the speed of sound. In this
work, we have used the algorithm proposed in Ref.
\cite{AhlrichsDuenweg99}, however, with the modification
that the original 18-velocity model (D3Q18) was replaced
by the D3Q19 19-velocity model \cite{Succi01}. The set of
$\mathbf{c}_i$ consists of the particle being at rest, the 6 nearest
and 12 next-nearest neighbors on a simple cubic lattice. The
magnitudes of the velocities corresponding to these three sets of
particles are $c_i = \left\vert \mathbf{c}_i \right\vert = 0,
a / \Delta \tau$, and $\sqrt{2} a / \Delta \tau$,
respectively. The weight factors for the D3Q19 model are $w_0 = 1/3$,
$w_1 = 1/18$ and $w_{\sqrt{2}} = 1/36$.

The early model of Ref. \cite{AhlrichsDuenweg99} only considered the
thermalization of modes related to the viscous stress tensor. It is
important to note that even though this procedure is correct in the
hydrodynamic limit, it provides poor thermalization on smaller
length scales \cite{Adhikarietal05}. Adhikari et al.
\cite{Adhikarietal05} have shown that by applying thermalization to
all nonconserved modes one gets a significantly improved numerical
behavior at short scales; the theoretical background is now
thoroughly understood \cite{Duenwegetal07,DuenwegLadd09}. In this
work, we have also investigated the effects of thermalization of the
kinetic modes on various dynamic properties.

The coupling to the beads is done via simple interpolation of the flow
velocity from the surrounding sites, and by introducing a
phenomenological Stokes friction coefficient $\zeta_{\text{bare}}$ of
the beads. This gives rise to a friction force on the particles, plus
a Langevin force that balances the frictional losses. The total
momentum is conserved by subtracting the corresponding momentum
transfer from the surrounding fluid. It can be shown that this
procedure satisfies the fluctuation-dissipation theorem
\cite{DuenwegLadd09}. For further technical details on this method and
its theoretical analysis, we refer the reader to
Ref. \cite{DuenwegLadd09}.

\subsection{Brownian dynamics (BD) simulations}

The configuration of a bead-spring chain is specified by the set of
position vectors $\mathbf{r}_{i}$ ($i = 1, 2, \ldots, N$). The time
evolution of this configuration is governed by the It\^{o} stochastic
differential equation \cite{OttingerBook96,PrabhakarPrakash04a,%
SuntharPrakash05}
\begin{eqnarray}
\nonumber
&&
\mathbf{r}_{i} (t + \Delta t) =
\mathbf{r}_{i} (t) + \left( k_{\text{B}} T \right)^{-1}
\mathbf{D}_{ij} \cdot
\left( \mathbf{F}_j^{\text{s}} +
       \mathbf{F}_j^{\text{int}} \right) \Delta t \\
&& +
\sqrt{2 \Delta t} \,
\mathbf{B}_{ij} \cdot \mathbf{W}_{j} ,
\quad \quad j = 1, 2, \ldots, N ,
\label{eq:SDE}
\end{eqnarray}
where summation over repeated indices is implied. Here the symbols
$\Delta t$, $k_{\text{B}}$, $T$, and $\mathbf{D}_{ij}$ denote the time
step, Boltzmann's constant, the temperature, and the diffusion tensor,
respectively, where the latter decribes the hydrodynamic interactions
between the beads. The forces $\mathbf{F}_{j}^{\text{s}}$ and
$\mathbf{F}_j^{\text{int}}$ are the spring and excluded-volume
contributions, repectively. $\mathbf{W}_{i}$ are random variables
representing a discretized Wiener process, such that $\left<
\mathbf{W}_{j} \right> = 0$ and $\left< \mathbf{W}_i \otimes
\mathbf{W}_j \right> = \mathbf{1} \delta_{ij}$, where $\delta_{ij}$ is
the Kronecker delta and $\mathbf{1}$ is the unit tensor. Finally, the
tensor $\mathbf{B}_{ij}$ is related to the diffusion tensor such that
$\mathbf{D}_{ij} = \mathbf{B}_{ik} \cdot \mathbf{B}_{jk}^{\text{T}}$
\cite{OttingerBook96}.

The frictional properties of the chain and the hydrodynamic
interactions between the beads are modeled via the diffusion tensor
$\mathbf{D}_{ij}$. Its diagonal elements contain the bead friction
coefficient $\zeta = 6 \pi \eta_s d$, where $\eta_s$ is the solvent
viscosity and $d$ is the Stokes radius of the bead. The off-diagonal
elements represent the hydrodynamic interactions via the tensor
$\boldsymbol{\Omega}_{ij}$, for which we take the regularized
Rotne-Prager-Yamakawa (RPY) tensor \cite{RotnePrager69,Yamakawa71}
with solvent viscosity $\eta_s$ and Stokes radius $d$. Taken together,
the diffusion tensor is given by
\begin{equation}
\frac{\mathbf{D}_{ij}}{k_{\text{B}} T} =
\zeta^{-1} \delta_{ij} \mathbf{1} + \left(1 - \delta_{ij} \right)
\boldsymbol{\Omega}_{ij} .
\end{equation}
Further details can be found in Ref. \cite{PrabhakarPrakash04a}.

The computationally most intensive part is to determine the matrix
$\mathbf{B}_{ij}$. Generally, Cholesky decomposition of
$\mathbf{D}_{ij}$ is used to obtain $\mathbf{B}_{ij}$ as an upper (or
lower) triangular matrix, and the computational cost for this method
scales as $N^3$ \cite{Fixman86}. Fixman made use of the fact that
there are many possibilities to define $\mathbf{B}_{ij}$ as some
square-root matrix of $\mathbf{D}_{ij}$, and, noting that it is the
vector $\mathbf{B}_{ij}\cdot \mathbf{W}_{j}$ that is required
rather than the matrix $\mathbf{B}_{ij}$, applied a truncated
Chebyshev polynomial expansion to obtain
$\mathbf{B}_{ij}\cdot \mathbf{W}_{j}$ with a lower computational
cost, scaling roughly as $N^{2.25}$ \cite{Fixman86}. In the present
paper, we make use of this accelerated technique as well.

\subsection{Unit systems and parameter mapping}

For implementation on a computer, physical quantities must be
represented in certain units, i.~e. in terms of suitable dimensionless
ratios. This is typically done by choosing a natural unit system where
three independent elementary quantities are set to unity.

In the coupled MD/LB simulation approach, this is usually done by
choosing the Lennard-Jones parameters $\epsilon$, $\sigma$, and $\tau$
($\tau = \sqrt{m \sigma^2 / \epsilon}$, where $m$ is the mass of the
monomer) as the units of energy, length, and time, respectively; this
choice enables one to make direct contact with MD simulations with
explicit solvent \cite{AhlrichsDuenweg99}. Conversely, BD simulations
have traditionally \cite{PrabhakarPrakash04a,SuntharPrakash05} used a
unit system where one chooses $k_{\text{B}} T$ as the energy unit and
$l_k = \sqrt{{k_{\text{B}}T}/{k_\text{FENE}}}$ as the length
unit. This is particularly useful for the simulation of pure Gaussian
(harmonic) chains where the interaction potential has neither an
energy scale nor a length scale built in. The time unit in BD simulations
is usually chosen as $\tau_k = \zeta / (4 k_\text{FENE})$.

Of course, a meaningful comparison of results requires that all data
are represented in one common unit system. At this point, one realizes
that this is less straightforward than one might think at first
glance. While the conversion of length and energy units is trivial,
and directly facilitated by the fact that both methods use the same
molecular model for the polymer chain, and perform the simulations at
the same temperature, the conversion of time units is not. This is so
because of the different time scales underlying the basic updating
algorithms: The MD/LB method is based upon simulating the system on
\emph{inertial} time scales, while BD focuses directly on the larger
\emph{diffusive} time scales. This is directly reflected by the
occurence of the intertial parameter $m$ (monomer mass) in the MD/LB
time unit, which does occur in the BD model, and the diffusive
parameter $\zeta$ (monomer friction constant) in the BD time unit. It
is important to notice that $\zeta$ is \emph{not} an input parameter
to the MD/LB model. Rather, one needs to carefully distinguish between
the short-time friction coefficient $\zeta_{\text{bare}}$ that is
indeed an input parameter --- it describes the Stokes coupling of the
monomer to the LB fluid in its immediate vicinity --- and the
long-time friction coefficient $\zeta_{\text{eff}}$ describing the
particle's long-time response that is modified by solvent backflow
effects. It is this latter parameter that must be identified with the
BD $\zeta$, and it is esentially an \emph{output} parameter.
Fortunately, the relation between $\zeta_{\text{bare}}$ and
$\zeta_{\text{eff}}$ is well understood --- Ahlrichs and D\"unweg
\cite{AhlrichsDuenweg99,DuenwegLadd09} have shown that
\begin{equation}
\label{eq:zeta_eff_vs_zeta_bare}
\frac{1}{\zeta_{\text{eff}}} =
\frac{1}{\zeta_{\text{bare}}} + \frac{1}{g \eta_s a} ,
\end{equation}
where $g$ is a numerical prefactor and $a$ is some measure for the
range of interpolation to the surrounding lattice sites. For linear
interpolation to the nearest sites, one finds $g \approx 25$, if $a$
is the LB lattice spacing. For highly accurate results, one should
also take into account a small correction for the finite size of the
simulation box \cite{DuenwegLadd09}; this has however not been done in
the present paper. Rather, we took the mapping determined in
Ref. \cite{AhlrichsDuenweg99} to calculate $\zeta_{\text{eff}}$ from
$\zeta_{\text{bare}}$ and identified this with the $\zeta$ parameter
of the BD calculations. Although the physical mapping done in this way
is essentially correct, it is important to notice that this aspect
introduces a certain amount of numerical inaccuracy when it comes to
quantitative comparisons.

Given the fact that it is intrinsically impossible to run the two
simulations with the same unit system, we chose to keep the previously
established systems of the two respective methods, and to map the
results \emph{a posteriori} by the procedure outlined
above. Furthermore, we chose to present all results in MD/LB
units. Technically, this means that for length and time unit
conversions we need to set $\bar{l} \sigma = l^* l_k$ and $\bar{t}
\tau = t^* \tau_k$, where the ``*'' superscript denotes BD
non-dimensionalization, while ``-'' denotes a non-dimensionalization
for MD/LB. The corresponding factors for the conversion from BD to
MD/LB units are then trivially found to be
\begin{eqnarray}
\frac{\sigma}{l_k} & = & \left( 
\frac{ \bar{k}_\text{FENE} \epsilon}{ k_\text{B} T }
\right)^{1/2} , \\
\frac{\tau}{\tau_k} & = & \frac{ 4 \bar{k}_\text{FENE} }{
\bar{\zeta}_\text{eff} } ;
\end{eqnarray}
note that $\bar{k}_\text{FENE}$ and $\bar{\zeta}_\text{eff}$ are
$k_\text{FENE}$ and $\zeta_\text{eff}$ in non-dimensional MD/LB units,
respectively, while the ratio $k_B T/ \epsilon$ is just the
non-dimensionalized temperature in MD/LB units.

The dimensionless hydrodynamic interaction parameter $h^*$ used in
the BD simulations is essentially a non-dimensionalized Stokes
radius. We thus find
\begin{equation}
h^* \sqrt{\pi} l_k = d = \frac{\zeta_\text{eff}}{6 \pi \eta_s} =
\frac{\bar{\zeta}_\text{eff} \sigma}{6 \pi \bar{\eta_s}} ,
\end{equation}
or
\begin{equation}
h^* = \frac{\bar{\zeta}_\text{eff}}{6 \pi^{3/2} \bar{\eta}} \left(
\frac{ \bar{k}_\text{FENE} \epsilon }{ k_\text{B} T} \right)^{1/2} .
\end{equation}

We therefore parameterized our simulations by first picking simulation
parameters for the MD/LB model (which were then directly used for
MD/LB), then converting these to BD units using the procedure outlined
above, and then running the thus-obtained equivalent BD model. For
these latter simulations, a time step size $\Delta t^* = 0.005$ (in BD
units) was found to produce sufficiently accurate results.

\subsection{Choice of parameters}

The physical input values for the present model are chosen from the
benchmark values developed in Ref. \cite{AhlrichsDuenweg99}, which
have been shown to reproduce the results of a typical pure MD
simulation \cite{DuenwegKremer93}. As in the comparison between LB and
MD simulations, we study a system of a single polymer chain of length
$N = 32$ monomers immersed in a fluid with temperature $k_\text{B} T /
\epsilon \ = 1.2$, density $\bar{\rho} = 0.864$, and kinematic
viscosity $\bar{\nu} = 2.8$. The lattice spacing $\bar{a}$ is set to
unity, which is roughly identical to the bond length; this is
necessary to resolve the hydrodynamic interactions on small length
scales with sufficient accuracy.

Furthermore, following Ref. \cite{AhlrichsDuenweg99}, the coupling
parameter $\bar{\zeta}_{\text{bare}}$ was set to $20.8$. The values of
the FENE spring potential parameters are $\bar{k}_\text{FENE}=7$ and
$\bar{R}_0=2$. The time step size for the polymer (the MD part of the
simulations) is set to $\Delta \bar{t} = 0.01$. It should be noted
that such a large time step is possible since the inclusion of
dissipation and noise lead to a substantial stabilization, compared to
purely microcanonical MD. The value of the time step that updates the
fluid should be chosen in a way such that the LB variables $n_i$ do
not become negative too often. Here, we choose $\Delta
\bar{\tau}=0.02$, where such a case rarely occurred during the
observation time. It is important to mention another free input
parameter which governs the time scale for the evolution of
hydrodynamic interactions, known as the Schmidt number $Sc = \nu /
D_0$, where $D_0$ is the diffusion constant of the single
monomer. This parameter can be set arbitrarily in the LB method by
choosing $\nu$ and $D_0$ (which can be tuned by choosing
$\zeta_\text{bare}$) accordingly. Ideally, the value of $Sc$ should be
chosen such that hydrodynamic interaction evolves much faster than the
diffusion of a monomer. In our case, we have $Sc \approx 32$, which
has been shown to result in Zimm-like behavior
\cite{AhlrichsDuenweg99,Jiangetal07}.

For the LB simulations, the polymer chain moves within a cubic box
of length $L$ with periodic boundary conditions, while it is
drifting freely in an infinite medium for the BD simulations. In
order to accurately compare various properties between the two
systems, one must understand the effects of the box length $L$ on
any observable of interest in the LB simulations. Thus it is
essential to only compare quantities under identical conditions
(i.~e. independent of the box length). Hence various box lengths $L$
ranging from 10 to 35 Lennard-Jones units were investigated; this
allows us to extrapolate to the $L \to \infty$ limit.

\section{\label{sec:Results} Results and discussion}
\subsection{Static properties}

The mean square radius of gyration and the mean square end-to-end
distance are given by
\begin{eqnarray}
\left< R^2_{\text{g}} \right> & = &
\frac{1}{2 N^2} \sum_{ij} \left< r^2_{ij} \right> , \\
\left< R^2_{\text{e}} \right> & = &
\left< \left( \mathbf{r}_{N} - \mathbf{r}_{1} \right)^2 \right>
\end{eqnarray}
with $r_{ij} = \left\vert \mathbf{r}_{i} - \mathbf{r}_{j} \right\vert$
being the inter-particle distance.

These two quantities are both related to the number of monomers by
the expression
\begin{equation}
\left< R^2_{\text{g}} \right> \propto
\left< R^2_{\text{e}} \right> \propto N^{2\nu} ,
\label{eq:Flory}
\end{equation}
where $\nu$ is the Flory exponent. For a self-avoiding walk (SAW),
the Flory exponent $\nu$ is 0.588 \cite{FloryBook88}. In principle,
$\nu$ can be obtained from simulations, using the scaling law in
Eq.~\ref{eq:Flory}. However, this method would require
simulations for a wide range of $N$ values. Alternatively, one can
use the static structure factor
\begin{eqnarray}
\nonumber
S(k) & = & \frac{1}{N}
\sum_{ij} \left<
\exp \left( i \mathbf{k} \cdot \mathbf{r}_{ij} \right) \right>
\\
& = & \frac{1}{N}
\sum_{ij} \left< \frac{\sin(kr_{ij})}{kr_{ij}} \right>
\end{eqnarray}
to obtain $\nu$ much more efficiently.

In the scaling regime $R^{-1}_\text{g} \ll k\ll a_0$ ($a_0$ being a
microscopic length of the order of the bond length), a power law
relation between the static structure factor and the wave vector
$k$ holds:
\begin{equation}
S(k) \propto k^{-1/\nu} .
\label{eq:Svsk}
\end{equation}

\begin{figure}[tbp]
\centerline{\includegraphics[width=1.0\textwidth]{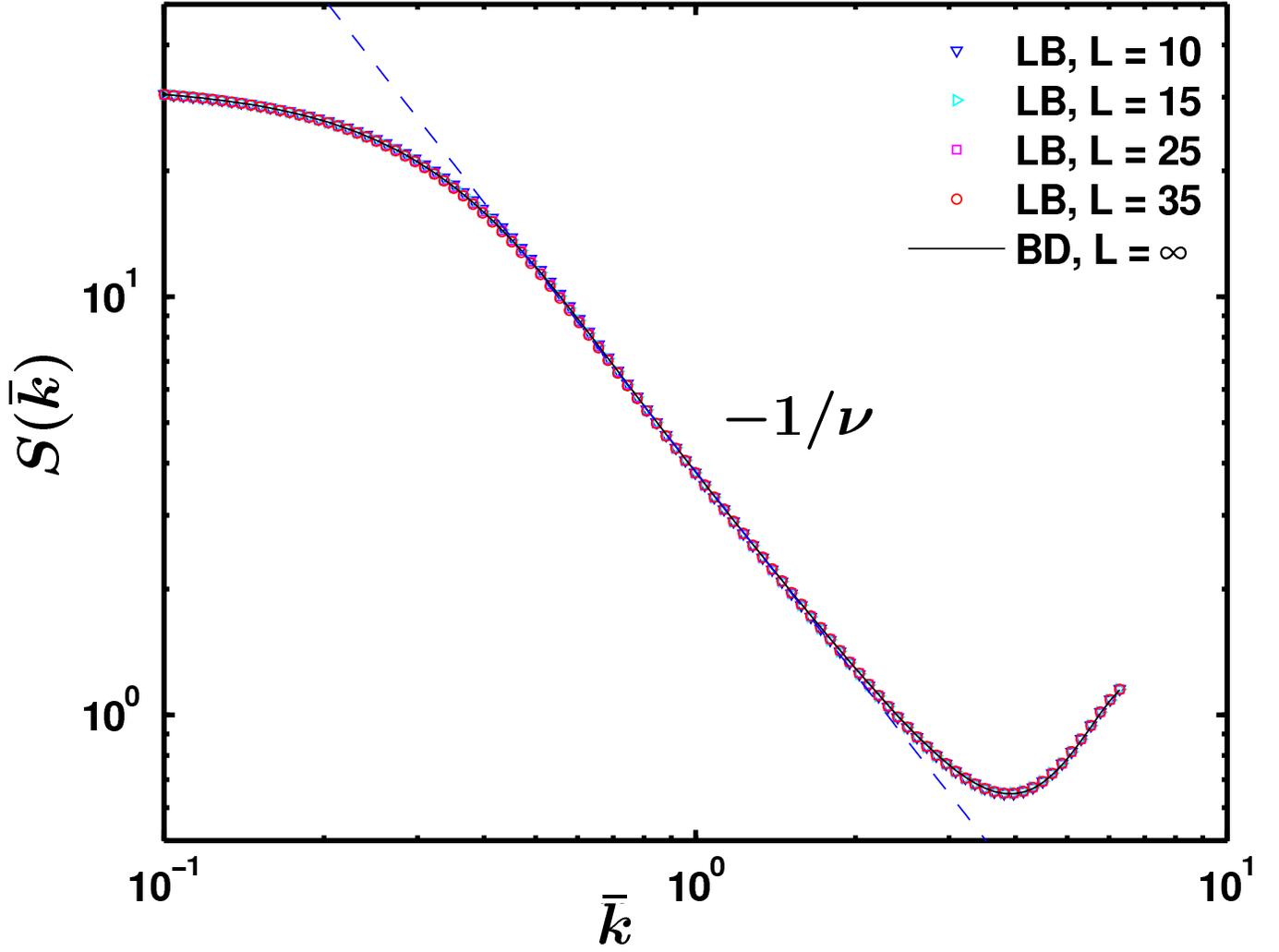}}
\caption{The static structure factor for the LB simulations (at
various box lengths $L$) and the BD simulations ($L=\infty$) for a
wide range of dimensionless wave vectors $\bar{k}$.}
\label{fig:Skstat}
\end{figure}

Figure~\ref{fig:Skstat} shows the static structure factor as a
function of wave vector $k$ for the LB simulations with the presence
of thermalization of all modes, and the BD simulations. It can be
clearly seen that the values of the static structure factor obtained
from the LB simulations are exactly the same as those obtained from
the BD simulations, indicating that they have the same static
conformations. From Eq.~\ref{eq:Svsk}, the value of $\nu$ can be
extracted from the linear region of the log-log plot of $S(k)$ vs $k$.
As expected, the values for $\nu$ obtained via this method are the
same for both the LB and the BD simulations, as reported in
Table~\ref{tab:Props}. However, they are approximately 5\% higher than
the asymptotically correct value, which is a consequence of the finite
chain length. The results for the mean square radius of gyration and
the mean square end-to-end distance in Table~\ref{tab:Props} further
confirm this agreement with regard to static conformations between the
two methods. However, at small box length ($L=10$), the results for
these static properties for the LB method deviate from their
asymptotic values. The discrepancy observed here always arises when
the box length is too small compared to the chain size, where the
chain is more likely to wrap over itself due to spatial restriction
and hence alter its static conformations. We also found that the two
versions of LB thermalization (``stresses--only'' vs ``full''
thermalization) yield identical results for the chain conformational
statistics. In general, we only quote values obtained for full
thermalization, unless indicated otherwise.

The hydrodynamic radius for a single chain in an infinite medium is
given by
\begin{equation}
\left< \frac{1}{R_\text{H}} \right>_\infty
= \frac{1}{N^2} \sum_{i \neq j} \left< \frac{1}{r_{ij}} \right> .
\end{equation}

For a chain in a finite box, as is the case here in the LB method, it
has been shown that the hydrodynamic interactions of the chain with
its periodic images effectively increases $R_\text{H}$
\cite{DuenwegKremer93,AhlrichsDuenweg99}. In order to account for this
finite-size effect, a finite-size correction of order $L^{-1}$ for
most dynamic properties, resulting from the slow $r^{-1}$ decay of
hydrodynamic interactions, is required
\cite{DuenwegKremer93,AhlrichsDuenweg99}.  The results for the
infinite-box value $\left< R^{-1}_{\text{H}} \right>_\infty$ agree
excellently with each other for all simulations (see
Table~\ref{tab:Props}). Since the overwrapping effect is more
sensitive to large inter-particle distances, it turns out that the
deviation in the inverse hydrodynamic radius is too small, for the
range of box lengths used, for it to be distinguishable. As can be
seen in Table~\ref{tab:Props}, the deviation is more pronounced for
the radius of gyration, and even more for the end-to-end distance.

\subsection{Dynamic properties}

According to dynamic scaling, the longest relaxation time
$\tau_{\text{Z}}$ of the chain is, by order of magnitude, identical to
the time that the chain needs to move its own size,
i.~e. $D_{\text{CM}} \tau_{\text{Z}} \sim R^2_{\text{g}}$, where
$D_\text{CM}$ is the diffusion constant of the chain's center of
mass. This leads to a dynamic scaling law $\tau_{\text{Z}} \propto
R^z_{\text{g}}$, where $z$ is the dynamic scaling exponent. For a
chain with hydrodynamic interactions, this relaxation time is known as
the Zimm time $\tau_{\text{Z}}$. For this case, $D_{\text{CM}} \propto
R^{-1}_{\text{g}}$ in the limit of long chains. This implies that
$\tau_{\text{Z}} \propto R^3_{\text{g}}$, which gives a dynamic
exponent of $z = 3$ for models with HI. For the Rouse model
(i.~e. chains without hydrodynamic interactions), where $D_{\text{CM}}
\propto N^{-1}$, one finds a dynamic exponent of $z = 2 +
1/\nu$. These quantities will be referred to in the discussion below.

\begin{figure}[tbp]
\centerline{\includegraphics[width=1.0\textwidth]{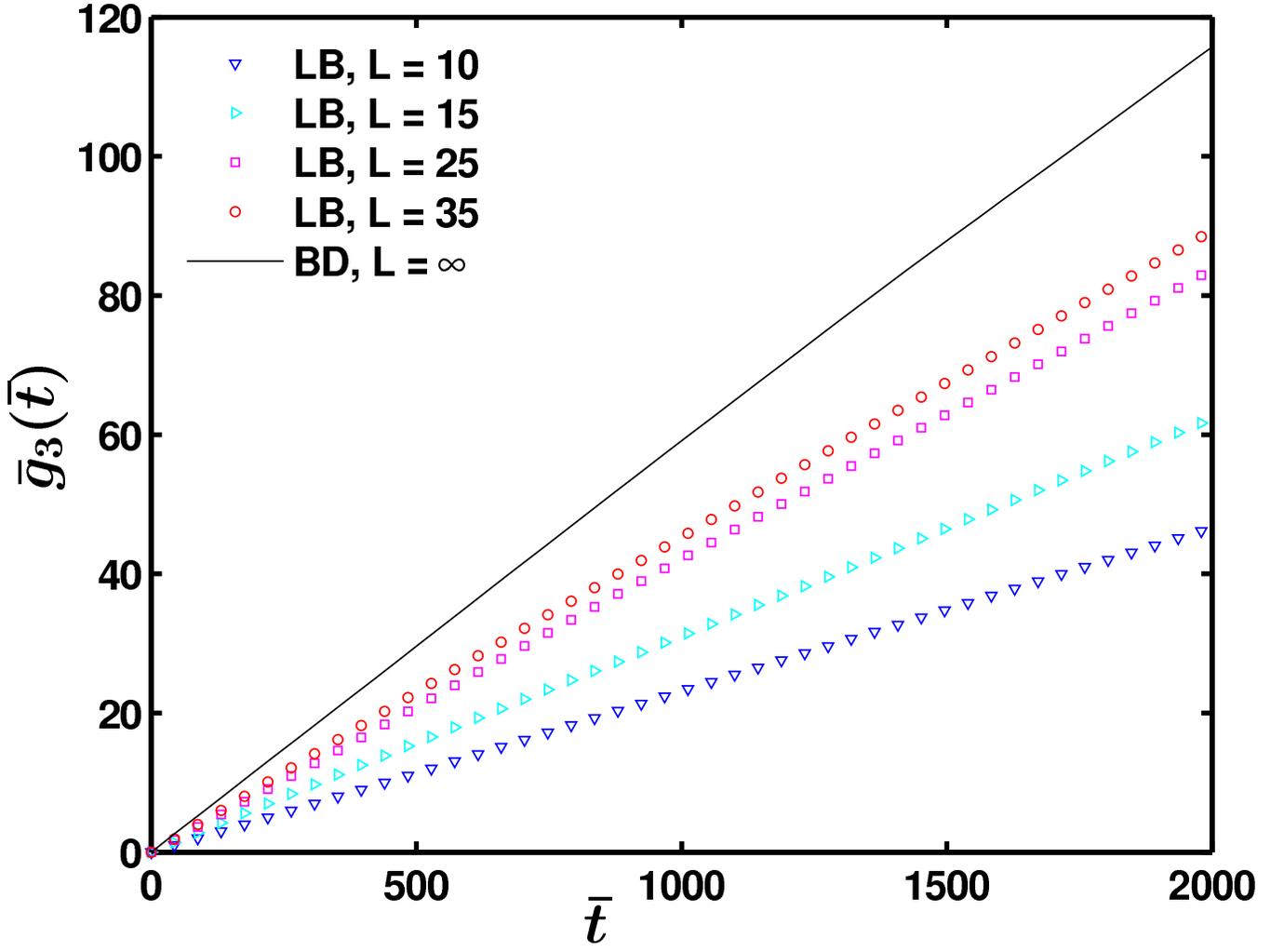}}
\caption{The dimensionless mean-square displacement
$\bar{g}_3(\bar{t})$ of the chain's center of mass,
Eq.~\ref{eq:g3}.} \label{fig:g3}
\end{figure}

The mean-square displacement of the chain's center of mass
\begin{equation}
g_3(t) = \left< \left( \mathbf{R}_{CM} (t_0 + t) -
                       \mathbf{R}_{CM} (t_0) \right)^2 \right>
\label{eq:g3}
\end{equation}
for both methods is depicted in Fig.~\ref{fig:g3}. From the figure, it
can be clearly seen that $g_3$ strongly depends on the box length $L$
for the LB simulations. Moreover, they seem to converge to the value
predicted by the BD simulations ($L=\infty$) in the limit of large
$L$. Effects of thermalization of the kinetic modes in LB simulations
on this property will be discussed subsequently. The chain's center of
mass diffusion constant $D_\text{CM}$ can be determined by the slope
of the $g_3$ vs $t$ curve, where the relationship $g_3(t) = 6
D_{\text{CM}} t$ holds. By fitting a power law to the simulation data,
we obtain the exponents and the diffusion constants shown in
Table~\ref{tab:Props}. These exponents support the prediction of
simple diffusive behavior ($t^1$). Theoretically, one would expect
that two diffusive regimes exist: On the one hand, there should be a
short-time diffusive regime, corresponding to time scales well below
the Zimm time, $t \ll \tau_{\text{Z}}$, but also well above the
ballistic regime, $t \gg \tau_0$; note that $\tau_0 > 0$ only in the
LB case, since the BD equation of motion is overdamped. On the other
hand, there should be free diffusion for times $t \gg
\tau_{\text{Z}}$. Both these regimes exhibit $t^1$ behavior, but with
different prefactors, with a smooth crossover around the Zimm time
\cite{LiuDuenweg03,Fixman81,Fixman83}. In principle these two
different diffusion constants can be obtained via fits to the
corresponding regimes. In practice, however, it turns out that the
values are very close to each other, and hence the crossover is very
smooth \cite{LiuDuenweg03,Fixman81,Fixman83}. Therefore its
unambiguous identification is very difficult, i.~e. impossible within
the resolution of our data.

\begin{figure}[tbp]
\centerline{\includegraphics[width=1.0\textwidth]{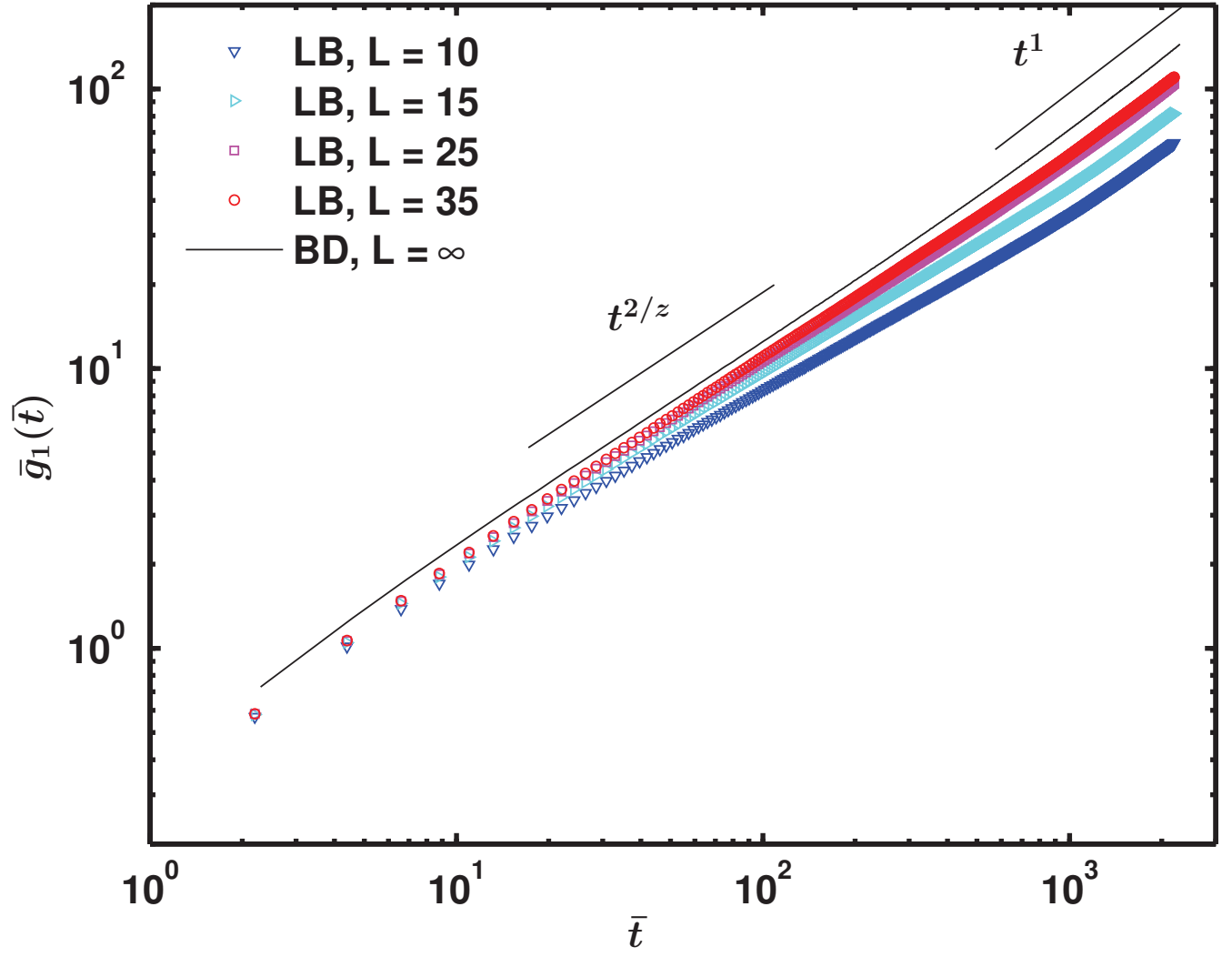}}
\caption{The dimensionless mean-square displacement
$\bar{g}_1(\bar{t})$ of the central monomer, Eq.~\ref{eq:g1}. Values
of the exponent $z$ at various box length $L$ in the sub-diffusive
scaling regime are also listed in Table~\ref{tab:Props}.}
\label{fig:g1}
\end{figure}

The mean-square displacement of a single monomer $i$ is given by
\begin{equation}
g_1(t) = \left< \left( \mathbf{r}_i(t_0 + t)
                      -\mathbf{r}_i(t_0) \right)^2 \right> .
\label{eq:g1}
\end{equation}
Here, only the two innermost monomers near the center of the chain
are evaluated to eliminate end effects; the results are plotted in
Fig.~\ref{fig:g1}. The values of $g_1$ behave similarly to those of
$g_3$. In the sub-diffusive time regime, corresponding to the
short-time diffusive regime for $g_3$, here evaluated between
$\bar{t} = 20$ and $80$, the scaling behavior $g_1(t) \propto
t^{2/z}$ is predicted \cite{DoiEdwards86}. The corresponding
exponents obtained from a power-law fit are listed in
Table~\ref{tab:Props} and indicate a value of $z=2.75$ as $L
\rightarrow \infty$. Regardless of the finite-size corrections due
to the box length and the effects of thermalization, these values
clearly favor the Zimm model compared to the Rouse model, which
predicts $g_1(t) \propto t^{0.54}$. The deviation from the
asymptotic Zimm value $z=3$ (or $g_1 \propto t^{2/3}$) is
mainly a result of finite chain length.

\begin{figure}[tbp]
\centerline{\includegraphics[width=1.0\textwidth]{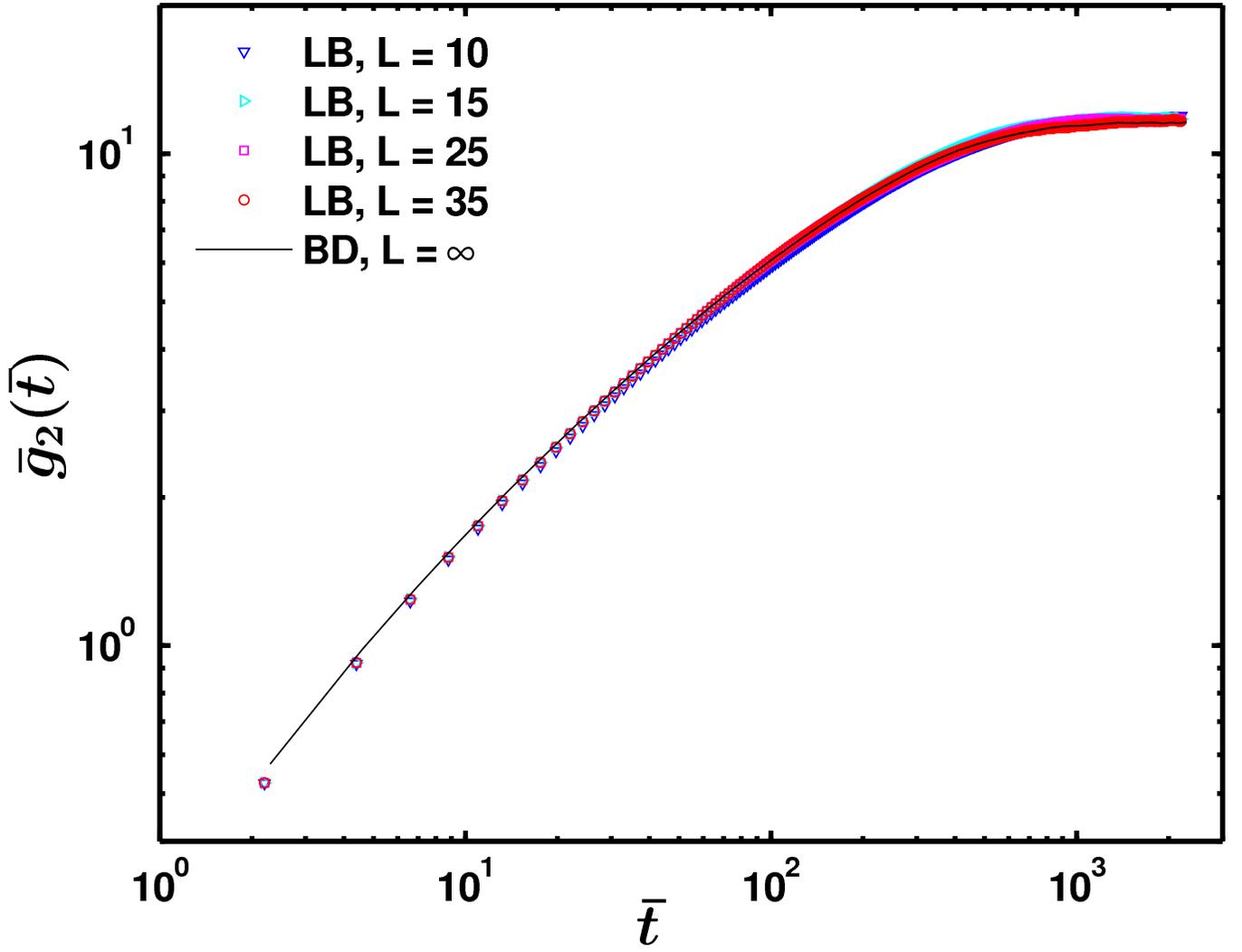}}
\caption{The dimensionless mean-square displacement
$\bar{g}_2(\bar{t})$ of the central monomer in the chain's center of
mass system, Eq.~\ref{eq:g2}.} \label{fig:g2}
\end{figure}

Figure~\ref{fig:g2} shows the mean-square displacement of a single
monomer in the center of mass system (i.~e. the two innermost
monomers to eliminate end effects)
\begin{eqnarray}
g_2(t) & = & \langle (
  \left[ \mathbf{r}_i(t_0 + t) - \mathbf{R}_\text{CM}(t_0 + t) \right] \\
& - & \left[ \mathbf{r}_i(t_0)     - \mathbf{R}_\text{CM}(t_0)
\right] )^2
      \rangle .
\nonumber
\label{eq:g2}
\end{eqnarray}

Interestingly, when viewed within the center of mass system, all the
results lie on top of each other, regardless of the box length $L$.
This result also holds for LB simulations without full thermalization.
This shows that the global center-of-mass motion of the chain is
actually the primary contribution to the deviations between LB and BD
results. In the data of Fig.~\ref{fig:g2} this contribution is
suppressed: In terms of Rouse modes, only the internal modes
remain. For these modes, however, it has been shown
\cite{AhlrichsDuenweg99} that the HI with the periodic images is much
weaker, while the leading-order $r^{-1}$ HI cancels out. Therefore,
the corresponding finite-size effect scales as $L^{-3}$ instead of
$L^{-1}$, and this is so small that it is invisible in
Fig.~\ref{fig:g2}.

Theoretically, these data can also be used for estimating the Zimm
time as the time where the crossover to the long-time plateau
occurs. However, the crossover is quite extended and smooth, making
it difficult to extract. We therefore estimated the Zimm time via
\begin{equation}
\tau_{\text{tr}} = \frac{\left< R^2_{\text{g}} \right>}{6
D_{\text{CM}}} . \label{eq:tauZ}
\end{equation}
Strictly speaking, this definition is only valid for a single chain
in an infinite medium, where there is no finite box size effect. In
the presence of finite box size, it becomes the definition for the
\emph{translational} time ($\tau_\text{tr}$) rather than the Zimm
time: The former is subject to an $L^{-1}$ finite-size effect, due
to the strong $L$-dependence of $D_{\text{CM}}$, while the latter,
being defined via the relaxation of internal modes, is only subject
to an $L^{-3}$ size effect, as discussed above. The translational
times obtained from Eq.~\ref{eq:tauZ} (as shown in
Table~\ref{tab:Props}) are indeed different for different box
lengths $L$, as expected. Conversely, the results displayed in
Fig.~\ref{fig:g2} indicate that the systems with different box sizes
have (essentially) all the same (internal-mode) Zimm time, since
their data all lie on top of each other.

\begin{figure}[tbp]
\centerline{\includegraphics[width=1.0\textwidth]{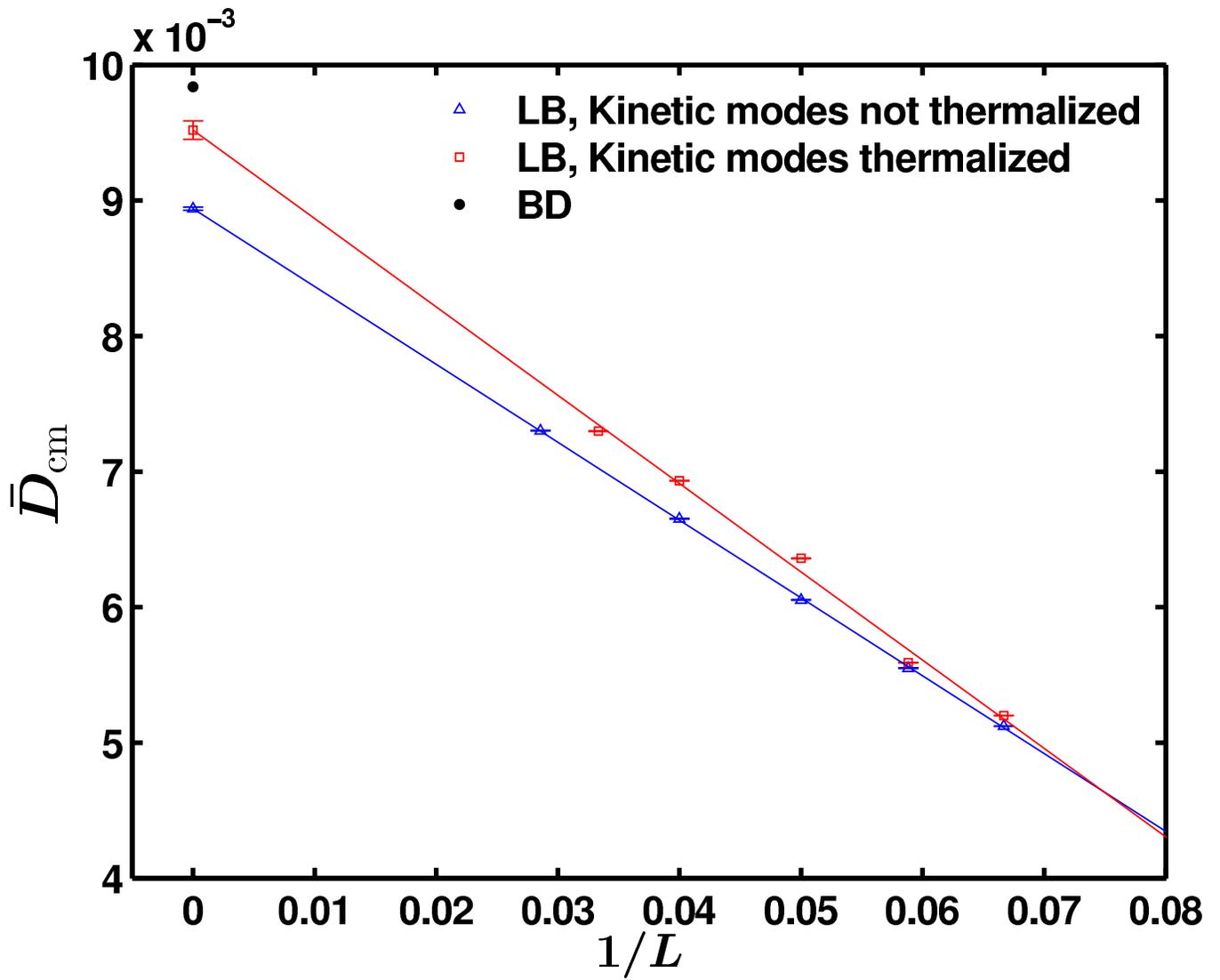}}
\caption{The dimensionless long time diffusion constant for the
center of mass at various box lengths $L$.} \label{fig:DcmvsL}
\end{figure}

Next, we focus on the leading order $L^{-1}$ finite size correction
for the long-time diffusion constant of the chain's center of mass,
$D_{\text{CM}}$. In principle, a plot of $D_{\text{CM}}$ vs $L^{-1}$
should give a straight line for large $L$, and an extrapolation to the
limit $L \to \infty$ should yield the same value as predicted by the
BD simulations. Figure~\ref{fig:DcmvsL} shows the values of
$D_{\text{CM}}$ for the LB simulations with and without thermalization
of the kinetic modes at various box lengths $L$ plotted together with
the value obtained from the BD simulation at $L = \infty$. It is worth
mentioning that the BD value of $D_{\text{CM}}$ can be obtained from
the mean square displacement of the chain center of mass or via
Fixman's expression \cite{Fixman81}. The latter method has been shown
to produce a much more reliable result and is easier to carry out
\cite{LiuDuenweg03}. The value reported here has been cross checked by
both methods and the results are almost the same within error bars.
For the LB simulations without thermalization of the kinetic modes,
the value of $D_{\text{CM}}$ at the asymptotic limit $L = \infty$ is
different from that predicted by the BD simulations by about $9.5\%$.
However, when all the kinetic modes in the LB simulations have been
thermalized, the deviation in $D_\text{CM}$ reduces to $3\%$. This
result clearly indicates that it is very important to thermalize all
the kinetic modes in order to obtain correct values for dynamic
properties.

The reason for the remaining small discrepancy between LB and BD is
not completely clear, since there are numerous possible sources.
Firstly, it should be noted that the underlying equations of motion
are quite different: LB works with inertia, while BD employs
overdamped dynamics. This results in different Schmidt numbers $Sc$
and different Mach numbers $Ma$, the latter being defined as the
ratio of the flow velocity to the speed of sound: Both are finite in
the LB method, while in the BD case they are strictly infinite
($Sc$) and zero ($Ma$), respectively.  Furthermore, the shape of the
HI function at small interparticle distances is somewhat different
for the two methods: In the BD case, we employ the RPY tensor, while
the nearest-neighbor interpolation for LB results in a short-range
HI that differs somewhat from the RPY tensor (see also the
discussion in Ref. \cite{DuenwegLadd09}).  Finally, it should be
noted that the value of the constant $g$ in Eq.
\ref{eq:zeta_eff_vs_zeta_bare}, which is crucial for the mapping
between the LB friction parameter $\zeta_{\text{bare}}$ and the BD
friction $\zeta_{\text{eff}}$, is only known with some numerical
inaccuracy. For highly accurate mappings, it is also necessary to
include a finite-size correction in the definition of $g$
\cite{DuenwegLadd09}; this was not done in the present study.

In order to examine whether the thermalization of the LB kinetic modes
is also important for the internal modes of chain motion, we have
performed a Rouse mode analysis. The Rouse modes for a discrete chain
are defined as \cite{Kopfetal97,AhlrichsDuenweg99}
\begin{equation}
\mathbf{X}_{p} =  \frac{1}{N} \sum_{n=1}^{N} \mathbf{r}_{n}
\cos  \left[ \frac{p \pi}{N} \left( n - \frac{1}{2} \right) \right]
\label{eq:Rmode}
\end{equation}
for $p = 1, 2, \ldots, N - 1$.

\begin{figure}[tbp]
\centerline{\includegraphics[width=1.0\textwidth]{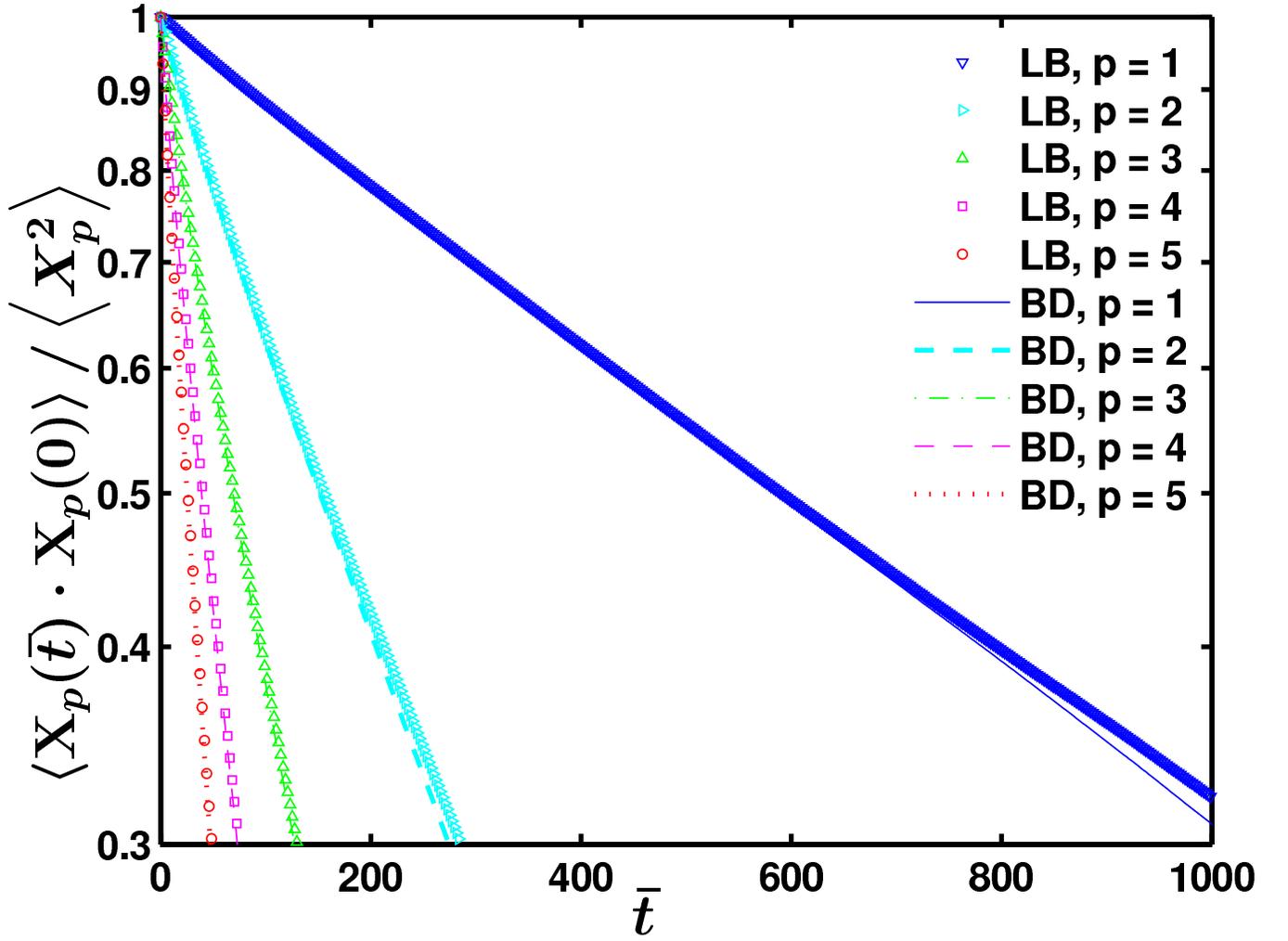}}
\caption{Normalized autocorrelation function of the first 5 Rouse
modes $\mathbf{X}_p$ (Eq.~\ref{eq:Rmodenorm}) for LB simulations at
$L=25$ and BD simulations at $L\rightarrow\infty$.}
\label{fig:Rmodeall}
\end{figure}

Within the approximation of the Zimm model, the autocorrelation
function of the modes should decay exponentially
\cite{DoiEdwards86}
\begin{equation}
\frac{ \left< \mathbf{X}_{p}(t_0 + t) \cdot \mathbf{X}_{p}(t) \right>}
{ \left< \mathbf{X}^2_p \right> } =
\exp \left( - \frac{t}{\tau_p} \right) ,
\label{eq:Rmodenorm}
\end{equation}
where $\tau_p$ is the relaxation of the $p$-th mode. To validate our
Rouse mode analysis routine, we have carried out extensive simulations
for a (Gaussian) Rouse chain of $N=8$ in the absence of HI and EV; the
results for $\tau_p$ are in excellent agreement with the analytical
predictions \cite{DoiEdwards86}. Figure~\ref{fig:Rmodeall} shows the
normalized autocorrelation function for $p = 1, 2, \ldots, 5$ for the
LB model with box length $L=25$, and the BD model. For nonzero times,
there is a small deviation between LB and BD, the latter exhibiting
again a slightly faster dynamics. This deviation systematically
becomes smaller upon increasing the mode index $p$. Since high mode
index means essentially relaxation on a rather small length scale, it
is tempting to attribute the deviation to the finite propagation of HI
in the LB model, i.~e. to retardation effects, or effects of finite
Schmidt number, which are more important on large length scales than
on small ones. Nevertheless, this hypothesis is not proven.

\begin{figure}[tbp]
\centerline{\includegraphics[width=1.0\textwidth]{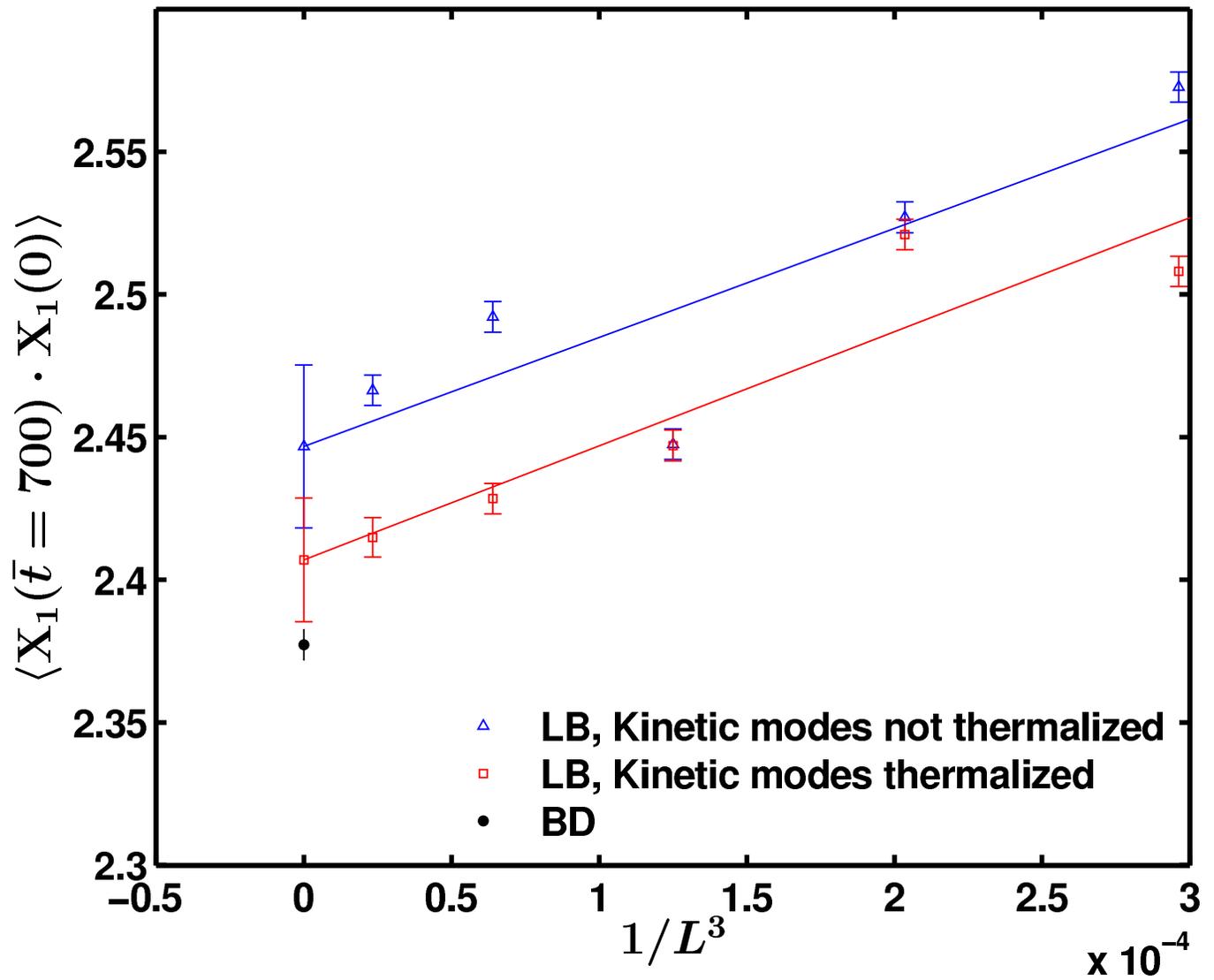}}
\caption{The autocorrelation function for the first Rouse mode
$\mathbf{X}_1$ at a finite time value of $\bar{t}=700$ for LB
simulations at various box lengths $L$ and BD simulations at
$L\rightarrow\infty$.} \label{fig:R1modevsL3}
\end{figure}

In Ref. \cite{AhlrichsDuenweg99} it was shown that the autocorrelation
function is only subject to an $L^{-3}$ finite size effect, in
contrast to the usual $L^{-1}$ behavior. Figure~\ref{fig:R1modevsL3}
shows the value of the autocorrelation function of the first Rouse
mode $\mathbf{X}_1$ at a fixed finite time $\bar{t}=700$, for LB
simulations at various box lengths $L$ and BD simulations at $L =
\infty$. Within our numerical resolution, the data indeed confirm this
$L^{-3}$ finite size effect, both with and without thermalization of
the kinetic modes. Furthermore, they demonstrate again that
thermalization of all the kinetic modes in LB simulations improves the
accuracy of the dynamic properties and brings them closer to the BD
prediction: The deviation in the extrapolated limit $L \to \infty$ is
reduced from $3\%$ down to $2\%$. The reasons for the remaining
discrepancies are probably of the same nature as in the case of
$D_{\text{CM}}$.


We have also evaluated the dynamic structure factor, which is
defined as
\begin{equation}
S(k,t) = \frac{1}{N} \sum_{ij} \left<
\exp \left( i \mathbf{k} \cdot
\left[ \mathbf{r}_i(t) - \mathbf{r}_j(0)  \right] \right) \right> .
\end{equation}

When both wave number and time are in the scaling regime (i.~e.
$R^{-1}_{\text{g}} \ll k \ll a^{-1}_0$ and $\tau_0 \ll t \ll
\tau_{\text{Z}}$), $S(k,t)$ is predicted \cite{DoiEdwards86} to
exhibit the scaling behavior
\begin{equation}
S(k,t) = S(k,0) f(k^zt) .
\end{equation}

\begin{figure}[tbp]
\centerline{\includegraphics[width=0.5\textwidth]{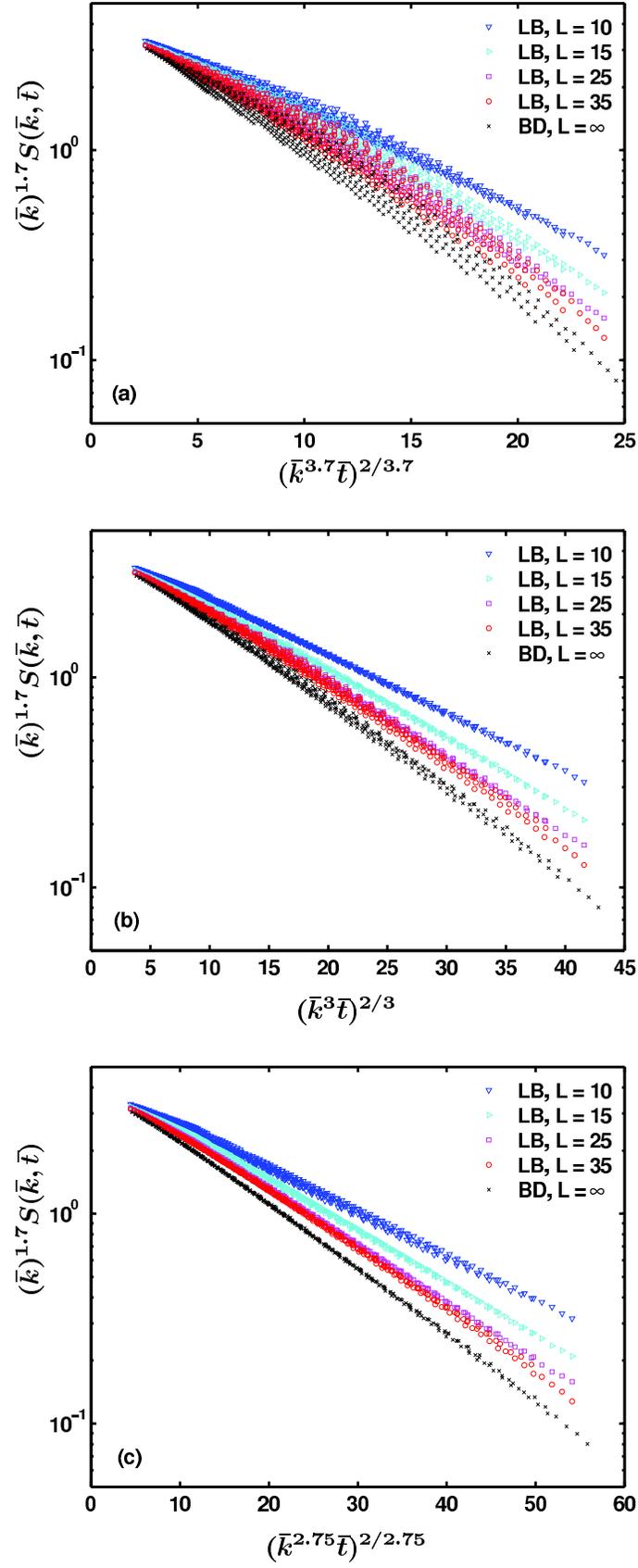}}
\caption{Scaling plot of the dynamic structure factor for (a) Rouse
scaling ($z=3.7$, top), (b) asymptotic Zimm scaling
($z=3$, center), and (c) $z=2.75$ (bottom),
which produces the best collapse.} \label{fig:Skz}
\end{figure}

A plot of $S(k,t) k^{1/\nu}$ against $(k^zt)^{2/z}$ should collapse to
a single curve \cite{AhlrichsDuenweg99}. The results for both methods
are shown in Fig.~\ref{fig:Skz}. The data were restricted to the
scaling regime $20 < t < 80$ and $0.7 < k < 1.5$. These ranges were
obtained from the single monomer mean-square displacement,
Fig.~\ref{fig:g2}, and from the static structure factor,
Fig.~\ref{fig:Skstat}, respectively. Here, we particularly focus on
adjusting the exponent $z$ such that it would produce the best total
data collapse for a chain in an infinite medium (i.~e. in the BD
model). Obviously, the results from the simulations show Zimm-like
rather than Rouse-like behavior. Even though we have suppressed the
finite box size effect, a dynamic exponent of $z=2.75$ yields the best
data collapse, which is somewhat smaller than the correct asymptotic
one. This result is also consistent with the value of $z$ obtained
earlier via the exponent of $g_1$ in the sub-diffusive scaling regime
(i.~e. $2/z=0.728$). The deviation from the asymptotic value is due
to the finite chain size used here, and one can expect $z=3$ only in
the long chain limit $N \rightarrow \infty$.

\begin{figure}[tbp]
\centerline{\includegraphics[width=1.0\textwidth]{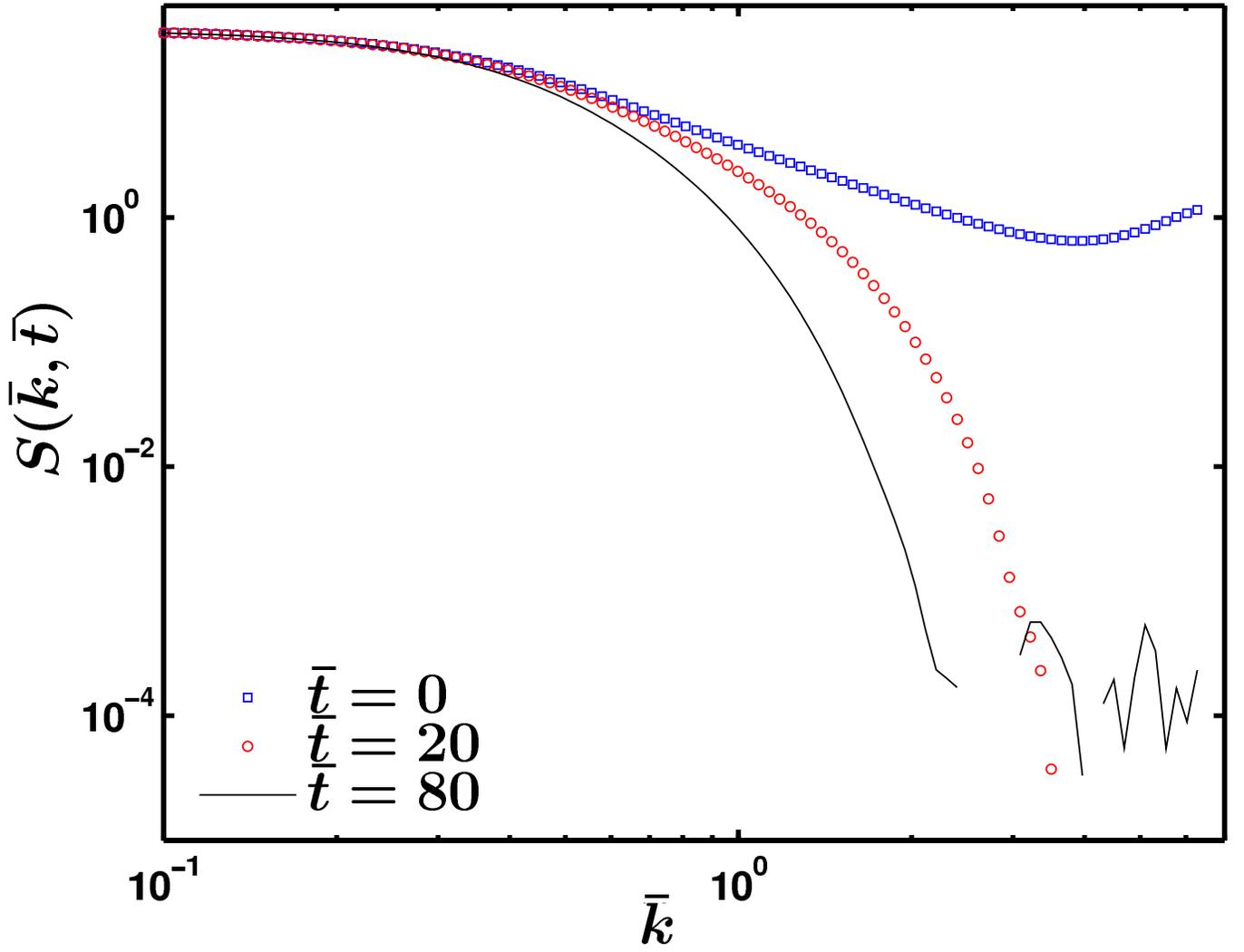}}
\caption{The dynamic structure factor $\bar{S}(\bar{k},\bar{t})$ for
the BD simulations ($L=\infty$) at three different times.}
\label{fig:SkallBD}
\end{figure}

\begin{figure}[tbp]
\centerline{\includegraphics[width=1.0\textwidth]{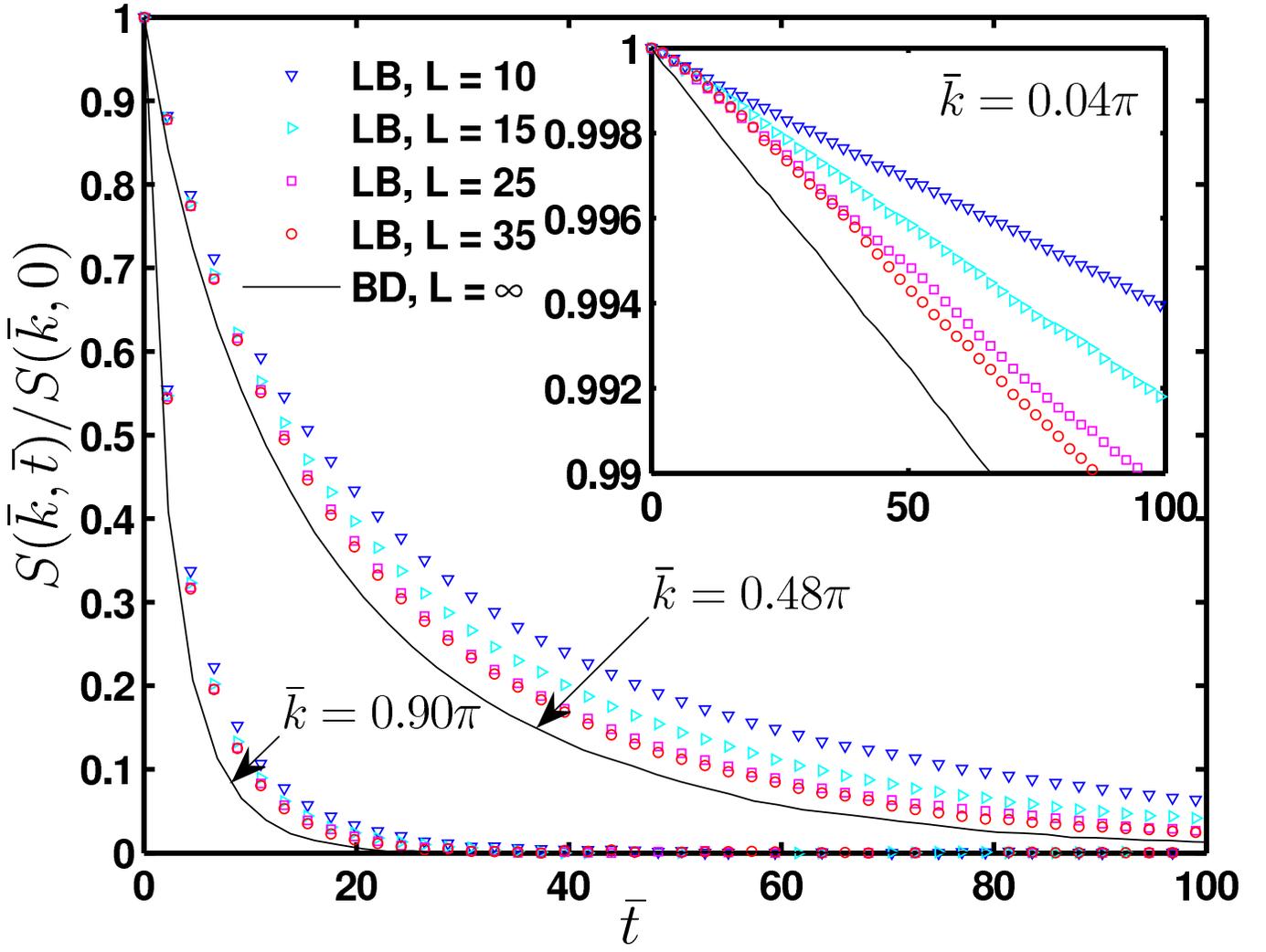}}
\caption{Time evolution of the normalized dynamic structure factor
at three different $\bar{k}$ values. Data for $\bar{k}=0.04\pi$ are
displayed in the inset.} \label{fig:StbySt0}
\end{figure}

More detailed comparisons of the structure factor
$S(\bar{k},\bar{t})$ are shown in Fig.~\ref{fig:SkallBD} ($\bar{k}$
dependence at constant time) and Fig.~\ref{fig:StbySt0} (time
dependence for the \textit{normalized} structure factor
$S(\bar{k},\bar{t})/S(\bar{k},0)$ at constant $\bar{k}$).

Figure~\ref{fig:SkallBD} shows the structure factor for BD
simulations for a wide range of $\bar{k}$ at three different times
and the data clearly indicate that the structure factor decays
rapidly with time. The normalized structure factor
$S(\bar{k},\bar{t})/S(\bar{k},0)$ for three different $\bar{k}$
values are shown in Fig.~\ref{fig:StbySt0}, and the data seem to
indicate that the LB results approach the BD data as $L$ is
increased, as expected.

\subsection{Efficiency}

\begin{figure}[tbp]
\centerline{\includegraphics[width=1.0\textwidth]{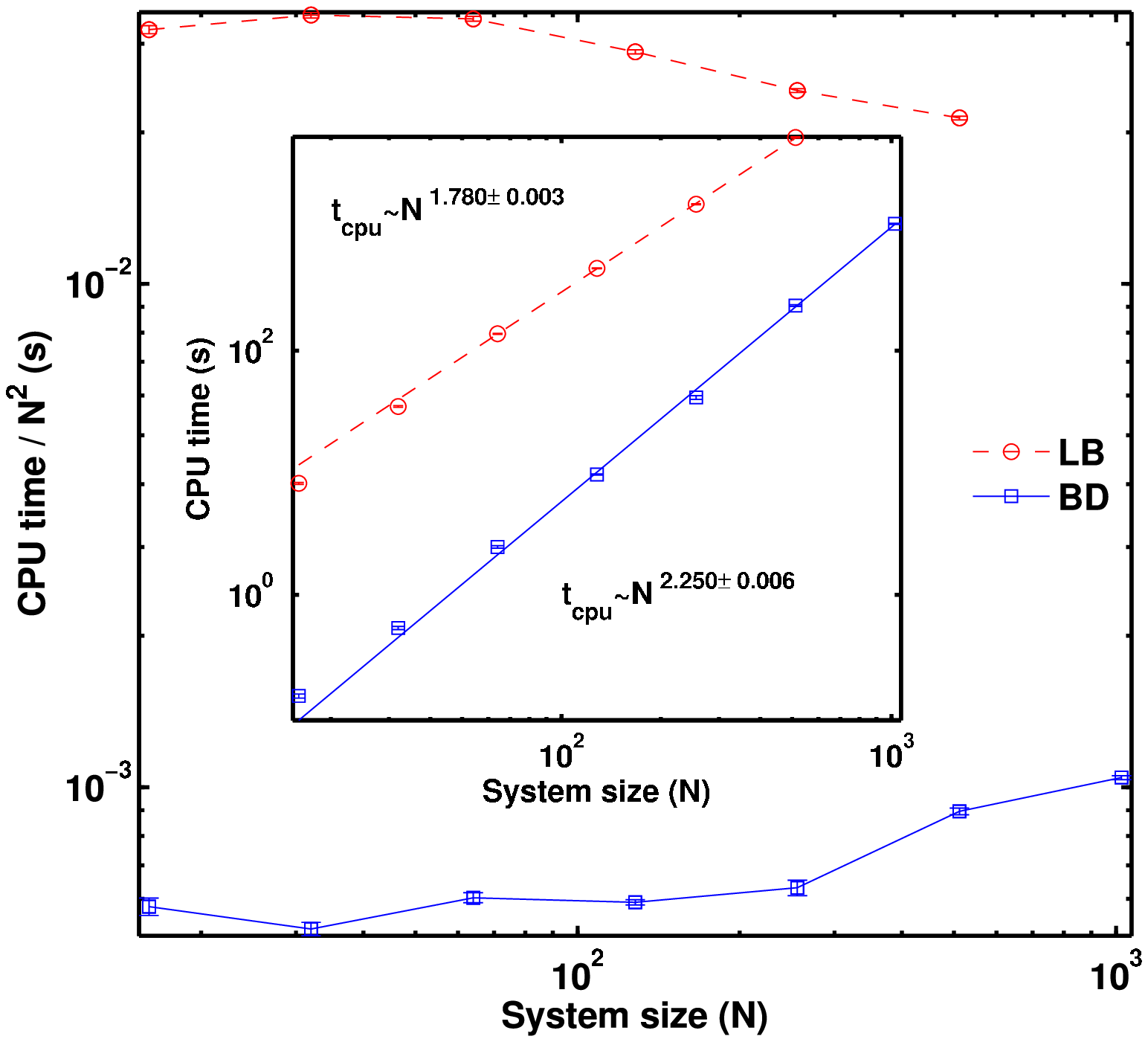}}
\caption{Comparison of the CPU time required by the LB and BD
systems for the equivalent of 1000 time steps for a wide range of
system sizes (chain lengths) $N$.} \label{fig:CPU}
\end{figure}

For the ultra-dilute system considered here, the lattice Boltzmann
part of the hybrid LB method uses up most of the computational
resources as the CPU cost for the MD part for the polymer chain is
negligible. Since the dynamic properties predicted by the LB model are
subject to a finite-size correction of order $L^{-1}$, extrapolation
is required to obtain these properties in the asymptotic limit $L
\rightarrow \infty$. To perform this extrapolation, together with
checking that indeed the asymptotic $L^{-1}$ behavior has been
reached, one needs the results of at least three different box
lengths. Moreover, the box length should be large enough compared to
the chain size such that it does not alter the static properties. The
data displayed in Table~\ref{tab:Props} indicate that it is safe to
choose $L$ such that $\sqrt{\left< R^2_{\text{e}} \right> } / L \leq
0.5$. The three different box lengths $L$ chosen here are
$\sqrt{\left< R^2_{\text{e}} \right> } / L = 0.5, 0.4$, and $0.3$. In
this work, we set the total CPU time required for the LB simulations
to be the sum of all the CPU times required to run 1000 MD time steps
for each of the chosen box lengths. For BD, we take the CPU time
needed to observe the system for the same time span in physical units.
Each of the simulations performed for the CPU time comparison was run
on an Itanium 2 processor of a 1.6 GHz SGI Altix server 3700. All the
parameters used to carry out this comparison are the optimal values
for both methods. Several chain sizes ranging from $N = 16$ to $1024$
have been used to obtain the CPU cost for comparison. The results are
shown in Fig.~\ref{fig:CPU}. For the LB method, it is clear that the
CPU cost scales linearly with the number of particles, i.~e. the
number of grid points that the solvent lives on, or $L^3$. Since the
ratio $\sqrt{\left< R^2_{\text{e}} \right> } / L$ is kept constant,
or $L \propto \sqrt{\left< R^2_{\text{e}} \right> } \propto N^{\nu}$,
this leads to a CPU cost scaling as $N^{3\nu}$.  This is indeed found
in our benchmarks, see Fig.~\ref{fig:CPU}.  Similarly, our data also
confirm the predicted $N^{2.25}$ CPU cost scaling for BD. Though the
LB exponent is lower than BD, the large prefactor ensures that the
total CPU cost for LB is much more expensive compared to BD for the
typical chain lengths used in the literature. It is only when the
chain length is excessively large (i.~e. $N$ of the order of $10^6$ or
higher) that LB will become superior to BD for a single-chain system.

The situation completely changes if one studies a semi-dilute solution
instead, as has been done in Ref.~\cite{ahlrichs:01}. For such a
system, we have not done a comparison between LB and BD in terms of
actual simulations; however, by means of scaling considerations one
can roughly estimate what the likely result of such a comparison would
be. A semidilute solution comprises $M$ chains of $N$ monomers each,
such that the total number of monomers is $M N$. Therefore the BD CPU
cost scales as $(M N)^{2.25}$, while the LB CPU cost depends on the
density. Within the blob picture of semidilute solutions, one views a
chain as a sequence of ``blobs'', each comprising $n$ monomers, and
having size $\xi$, which can be viewed as the typical correlation
length of density fluctuations, or the typical distance from which
point on chain-chain interactions become important. Since the
conformation statistics within the blob is that of a self-avoiding
walk, one has $\xi \sim a n^\nu$, where $a$ is the monomer size. The
sequence of blobs forms a random walk, hence $R_e \sim \xi
(N/n)^{1/2}$. This gives the minimum size of the simulation box, i.~e.
$L \sim \xi (N/n)^{1/2} \sim a n^\nu (N/n)^{1/2} = a N^\nu (n/N)^{\nu
  - 1/2}$, or $L^3 \sim a^3 N^{3\nu} (n/N)^{3\nu - 3/2}$. We thus see
that the CPU effort for the LB method is even slightly
\emph{decreased} by the factor $(n/N)^{3\nu - 3/2}$ compared to the
single-chain case at the same $N$, due to the shrinkage of the chains
resulting from excluded-volume screening. In order to estimate the
number of chains $M$, we note that the arrangement of blobs is
space-filling, i.~e. $L^3 \sim \xi^3 M (N/n) \sim a^3 M n^{3\nu}
(N/n)$. Comparing this with the previous expression for $L^3$, one
finds $M \sim (N/n)^{1/2}$. Therefore the BD effort, compared to the
single-chain case, is increased by a factor of $M^{2.25} \sim
(N/n)^{1.125}$. Taken together, this means that the ratio between LB
effort and BD effort is changed by a factor of $\sim (N/n)^{3\nu +
  1.125 - 1.5} \approx (N/n)^{1.425}$ in favor of LB. For $N/n = 30$,
which is needed as a minimum to resolve the Gaussian statistics of
the chains as a whole, one obtains a factor of $130$, which more or
less compensates the two orders of magnitude seen in
Fig.~\ref{fig:CPU}. Taking into account that for such a system the
BD simulation would have to calculate the HI with the periodic
images, e.~g., via Ewald sums, which is much more complicated than
the present single-chain simulation, one sees that for a semidilute
solution clearly LB is more efficient, unless a ``superfast'' BD
algorithm
\cite{BanchioBrady03,Saintillanetal05,Hernandez-Ortizetal07} is
used. For the latter case, the answer is not yet known. The results
of Ref. \cite{Chenetal07} indicate that LB/MD may be favorable for a
rather small number of monomers; however, this study was done (i) in
a non-trivial geometry, which implies a more complicated BD method,
and (ii) under complete neglect of thermal fluctuations in the LB
simulations, resulting in a substantial reduction in CPU effort.
Such a (partial or complete) neglect of thermal noise is sometimes
justified in strong nonequilibrium situations such as studied in
Ref. \cite{Chenetal07}; in that particular case, the justification
was checked by additional tests \cite{ChenPC}. Another possible
situation where LB noise is negligible is the case of strong
coarse-graining, where a single lattice site can already be
considered as a macroscopic thermodynamic system (for a detailed
discussion, see Ref. \cite{DuenwegLadd09}). However, in the general
case, and certainly in thermal equilibrium or weak nonequilibrium,
the proper inclusion of thermal noise is necessary, as demonstrated
theoretically in detail in Ref. \cite{DuenwegLadd09}, and also
corroborated by the present numerical results. For the general case,
the estimate of the LB CPU effort given in Ref. \cite{Chenetal07} is
therefore too optimistic.

\section{\label{sec:Conclusions}Conclusions}

The present study has shown that Brownian dynamics simulations are
capable of reproducing various properties predicted by a hybrid LB/MD
model (or vice versa). We have demonstrated how to obtain the input
values for the BD simulations from the physical input parameters of
the LB model such that both models would produce the same static and
dynamic properties. For the LB model, most dynamic properties are
subject to a finite-size correction of order $L^{-1}$. In addition to
this, it is very important to thermalize all the kinetic modes in
order to obtain the correct dynamic properties. Those results that are
not affected by $L^{-1}$ finite size effects, such as the mean square
displacement in the center of mass system, or the Rouse mode
autocorrelation function, agree very favorably with each other. For
highly dilute systems where the simulation of a single chain is
sufficient, BD is usually the method of choice, as it is much more
efficient than the coupled LB/MD approach, and finite box size
effects are absent. The situation changes however in the semidilute
case, where it is easy to estimate that BD will not be able to
compete, unless ``superfast'' algorithms are used. Moreover, one
should take into account that the hybrid LB/MD algorithm is rather
easily adaptable to complicated boundary conditions, and can even be
applied to flows at high Reynolds numbers, where the fluid degrees of
freedom become \emph{intrinsically} important, and cannot be handled
in terms of a Green's function.

\begin{acknowledgments}
This work was supported by the Australian Research Council under the
Discovery Projects program and the Mainz Max Planck International
Research School. Computational resources were provided by the
Australian Partnership for Advanced Computation (APAC) and the
Victorian Partnership for Advanced Computation (VPAC). Stimulating
discussions with A. J. C. Ladd are gratefully acknowledged.
\end{acknowledgments}

\widetext

\clearpage

\begin{table}[tbp]

\begin{tabular}{|l|c|c|c|c|}
\hline & \multicolumn{3}{|c|}{LB} & BD\\\hline
Box length $L$ & 10 & 15 & 25 & $\infty$\\
Time step & 0.02 & 0.02 & 0.02 & 0.005\\\hline
exponent $\nu$ & 0.615 $\pm$ \, 0.005 & \, 0.617 $\pm$ 0.005 \, & \, 0.619 $\pm$ 0.005 \, & 0.619 $\pm$ 0.004 \, \\
$\left< \bar{R}^2_{\text{e}} \right>$ & 94.56 $\pm$ 1.20 \, \, & 100.05 $\pm$ 1.26 \, \, & 100.20 $\pm$ 1.28 \, \, & 99.22 $\pm$ 1.24  \, \,\\
$\left< \bar{R}^2_{\text{g}} \right>$ & 14.83 $\pm$ 0.10 \, \, &
\, 15.31 $\pm$ 0.11 \, \, & \, 15.36 $\pm$ 0.11 \, \, & 15.25 $\pm$ 0.11 \, \,\\
$\left< \bar{R}_{\text{H}}^{-1} \right>_{\infty}$ & 0.291
$\pm$ 0.0005 & \, 0.290 $\pm$ 0.0005 & \, 0.289 $\pm$ 0.0005 & 0.290 $\pm$ 0.0005\\
$\bar{g}_1\text{-exp.}^{\text{a}}$ & 0.640 $\pm$ 0.0005 & \, 0.675 $\pm$ 0.0005 & \, 0.710 $\pm$ 0.0005 & 0.728 $\pm$ 0.0006\\
$\bar{g}_1\text{-exp.}^{\text{a,b}}$ & 0.645 $\pm$ 0.0006 & \, 0.684 $\pm$ 0.0006 & \, 0.714 $\pm$ 0.0006 & 0.728 $\pm$ 0.0006\\
$\bar{g}_3\text{-exp.}^{\text{a}}$ & 1.008 $\pm$ 0.0008 & \, 1.020 $\pm$ 0.0008 & \, 1.050 $\pm$ 0.0008 & 0.995 $\pm$ 0.0008\\
$\bar{D}_{\text{CM}}\times 10^{-3}$ & 3.914 & 5.162 & 6.959 & 9.843\\
 & $\pm 1 \times 10^{-3}$ & $\pm 1 \times 10^{-3}$ & $\pm 2 \times 10^{-3}$ & $\pm 1 \times 10^{-2}$ \\
$\bar{\tau}_\text{tr} (\text{estimate})$ & 631.36 $\pm$ 4.43 &
492.01 $\pm$ 3.56 & 368.51 $\pm$ 2.67 & 258.28 $\pm$ 1.87\\\hline
\end{tabular}

\caption{Properties for a single chain of length $N=32$ obtained
from lattice Boltzmann simulations at various finite box lengths and
Brownian dynamics simulations in infinite medium.
$^{\text{a}}$ Exponent obtained by fitting a power law in the
sub-diffusive scaling regime of the chain in Lattice Boltzmann
simulations, $\bar{t} \in \left[ 20:80 \right]$.
$^{\text{b}}$ Exponent obtained from Lattice Boltzmann simulations
without thermalization of all the kinetic modes.
}

\label{tab:Props}

\end{table}

\printtables
\clearpage
\printfigures

\end{document}